\documentclass[aps,pre,twocolumn,groupedaddress,floatfix]{revtex4-2}
\usepackage[english]{babel}

\usepackage{graphicx}
\usepackage{amsmath,amssymb,amsfonts}
\usepackage{dcolumn}
\usepackage{bm}

\begin{document}


\title{Pattern Formation in Coiling of Falling Viscous Threads: \\Revisiting the Geometric Model}



\author{Will Sze}
    \affiliation{Gina Cody School of Engineering and Computer Science, Concordia University, Montreal, QC, Canada}
\author{Eusebius J. Doedel}
\author{Ida Karimfazli}
    \email[]{ida.karimfazli@concordia.ca}
    \affiliation{Gina Cody School of Engineering and Computer Science, Concordia University, Montreal, QC, Canada}
\author{Behrooz Yousefzadeh}
    \email[]{behrooz.yousefzadeh@concordia.ca}
    \affiliation{Gina Cody School of Engineering and Computer Science, Concordia University, Montreal, QC, Canada}

\date{\today}

\begin{abstract}
    The ``Fluid Mechanic Sewing Machine'' creates periodic patterns through the coiling nature of a viscous fluid falling onto a moving surface. At relatively moderate heights, the reported patterns are translating coiling, alternating loops, W pattern, and meander. A simplified theoretical model based on the geometry and local bending of the contact point can predict these patterns. We experimentally explore the patterns in this region by collecting new data to compare with the model. Our review of the model's bifurcation diagram reveals additional patterns beyond the ones reported, although current experiments have not shown their existence. The W pattern, previously omitted in a regime diagram because of its small region, is now shown explicitly. We report on the consistent appearance of a period-doubled version of the W pattern, as well as rare appearances of resonant patterns, both reported for the first time. Comparing the theoretical model to experimental data, we find that the predicted phase diagram and the meander variation deviate from observations. These deviations hint at an unaccounted dynamics that merits further study.
\end{abstract}


\maketitle

\section{\label{Intro}INTRODUCTION}
It is well established in solid mechanics that thin features can buckle under sufficiently large compressive loads. Similarly, when viscous fluids are stretched to form thin threads or sheets, they can also exhibit buckling-like behaviours. When falling on a stationary surface, thin sheets periodically fold while thin threads coil. This spiralling behaviour is known as ``liquid rope coil" (LRC) and its dynamics have been extensively studied. An overview of the topic is discussed in  \cite{ribe_LRCOverview_2012,ribe_LRCSynoptic_2017}. The corresponding coiling frequency is understood to behave under four distinct regimes: viscous ($\Bar{V}$), gravitational ($\Bar{G}$), inertial-gravitational ($\Bar{IG}$) and inertial ($\Bar{I}$) \cite{ribe_coiling_2004,ribe_IG_2006}.

Changing the lower stationary boundary to a moving surface unravels the coils into periodic patterns likened to stitching, hence the name ``Fluid Mechanic Sewing Machine" (FMSM) \cite{chiu-webster_fall_2006}. 
In experiments, the varying parameters are the fluid fall height $\hat{H}$ and the moving surface speed $\hat{V}$, while parameters such as the volumetric flow rate $\hat{Q}$ and the nozzle diameter $\hat{d}$ are kept constant. The experiment is performed for a given fluid with kinematic viscosity $\hat{\nu}$, density $\hat{\rho}$, and surface tension $\hat{\sigma}$. 

\begin{figure}[!b]
	\centerline
        {\includegraphics[scale=0.25]{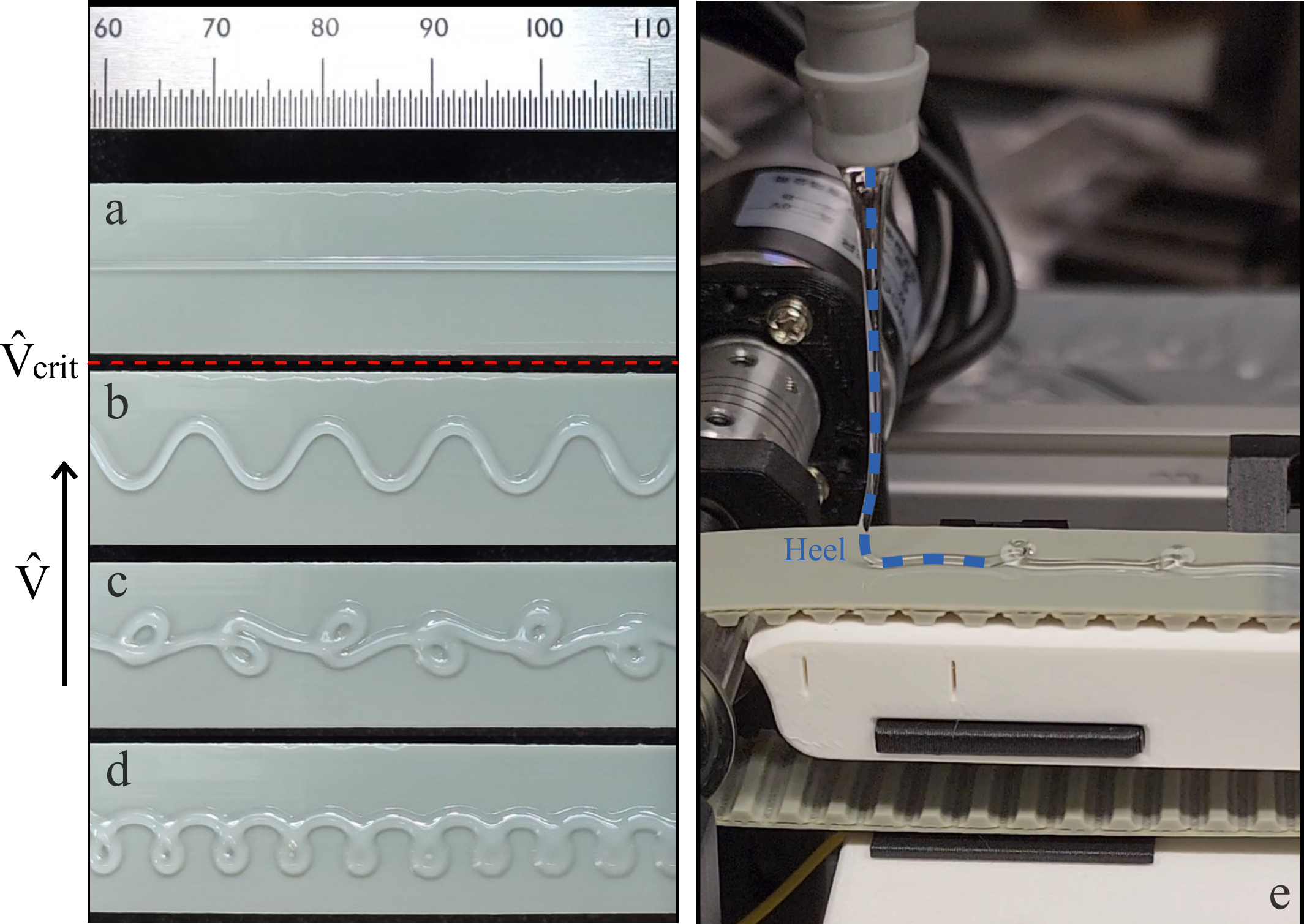}}
	\caption{(Color online) Belt patterns appearing at relatively low to moderate height levels. Patterns a-d are catenary (straight), meander, alternating loops and translating coiling, respectively. The arrow on the left indicates the order of increasing $\hat{V}$ with the critical speed $V_{crit}$ indicated by the dotted red line. On the right (e), the less common W pattern is shown. The dashed blue line highlights the heel of the thread.}
	\label{fig Patterns Intro}
    \end{figure}

The typical sequence of the ensuing patterns is as follows. At large $\hat{V}$, the thread is stretched, tracing a straight line (Fig.~\ref{fig Patterns Intro} a). As $\hat{V}$ is reduced to some critical speed $\hat{V}_{crit}$, the thread forms a ``heel-shaped" curve, indicating compression in the thread near the moving surface (Fig.~\ref{fig Patterns Intro} e).
The thread meanders when $\hat{V}$ is decreased below $\hat{V}_{crit}$  (Fig.~\ref{fig Patterns Intro} b). Further reducing $\hat{V}$ leads to numerous looping patterns (Fig.~\ref{fig Patterns Intro} c \& d). Depending on the values of $\hat{H}$, the sequence of patterns might be well-defined (at low to moderate heights) or seemingly unorganized and exhibiting large hysteresis (at higher heights). 
These two categories of patterns (low and high fall heights) correspond to the $\bar{G}$ and $\bar{IG}$ regimes of the LRC \cite{morris_meandering_2008,brun_DVT_2012}. 

The current paper focuses on the patterns appearing at moderate height levels. In this range, three patterns, herein called basic patterns, have been experimentally reported \cite{chiu-webster_fall_2006,morris_meandering_2008,welch_frequency_2012}. These are named translating coiling (Fig.~\ref{fig Patterns Intro} b), alternating loops (Fig.~\ref{fig Patterns Intro} c) and meander (Fig.~\ref{fig Patterns Intro} d).

Many have studied this phenomenon in detail \cite{chiu-webster_fall_2006,morris_meandering_2008,welch_frequency_2012,ribe_stability_2006,BLOUNT_LISTER_2011,brun_DVT_2012,brun_GM_2015}. Chiu-Webster and Lister \cite{chiu-webster_fall_2006} pioneered this study by conducting various experiments and collecting a plethora of patterns. At moderate heights, the three basic patterns appeared. Soon after, the experimental apparatus and data collection were substantially improved by Morris et al. \cite{morris_meandering_2008} and Welch et al. \cite{welch_frequency_2012}, adding more detail to the regime diagrams in the $\hat{V}$-$\hat{H}$ space. In addition to those three patterns, Welch et al. \cite{welch_frequency_2012} observed the appearance of the W pattern (Fig.~\ref{fig Patterns Intro} e), although its occurrence was rare and not explicitly shown in their regime diagram. 

Theoretical and numerical models were later developed to analyze and predict the experimentally observed patterns. Ribe et al. \cite{ribe_stability_2006} performed linear stability analysis to numerically predict $\hat{V}_{crit}$ and the frequency at which the thread starts to meander when approached from the catenary (straight) pattern. Later, Brun et al. \cite{brun_DVT_2012} numerically reproduced many of the patterns seen experimentally using the Discrete Viscous Threads (DVT) algorithm. This computational method greatly simplifies the modelling of thin features by condensing their properties onto the centerline \cite{Bergou_DVT_2010}. A more simplified theoretical approach developed by Brun et al. \cite{brun_GM_2015} captures the basic patterns observed at the lower height ranges. This reduced model, named the Geometrical Model (GM), is a set of three Ordinary Differential Equations (ODEs) with the position of the contact point and the tangent of the thread curvature as state variables. The GM is developed based on the no-slip boundary condition at the contact point and does not account for the inertia of the falling thread. The earlier results from DVT had confirmed that inertial effects are not significant at these parameter ranges~\cite{brun_GM_2015}. Interestingly, however, the GM captures not only the non-inertial patterns formed by a Newtonian fluid thread but also by other materials such as non-Newtonian visco-elastic fluids~\cite{yuk_new_2018}, yield-stress fluids~\cite{geffrault_printing_2023}, and elastic materials \cite{habibi_pattern_2011,jawed_geometric_2015}. 

The stability analysis of Ribe et al.~\cite{ribe_stability_2006} correctly predicts the critical belt speeds when the straight pattern transitions to meandering, along with the resulting oscillations.
The results of DVT and GM have been shown to agree well with experimental observations.  
Using DVT, Brun et al. \cite{brun_DVT_2012} demonstrated that, at small heights, the numerical model accurately replicates the pattern regions down to their boundaries (transition points). Using GM, Brun et al. \cite{brun_GM_2015} compared the transitions between the patterns with the DVT, indicating some similarities. They also showed that the W pattern can appear at low to moderate heights while increasing the surface speed. The comparisons between numerical and experimental studies were mostly focused on the regions and boundaries appearing in the $\hat{H}$-$\hat{V}$ regime diagram. Although some studies have shown how a specific pattern varies for a given parameter \cite{brun_DVT_2012,welch_frequency_2012}, to our knowledge, no direct comparisons between numerical and experimental results are made regarding these variations.

In this work, we conduct new experiments and re-examine the patterns that can appear from a viscous thread falling onto a moving surface. We compare the patterns observed in the newly gathered experimental data and numerical results. We use quantitative measures to characterize and compare the patterns. We focus on patterns appearing at relatively low to moderate heights. Accordingly, we use the GM for comparison with experimental data. 
Considering the model's ability to predict the basic patterns, it will be intriguing to determine whether it can accurately predict the variations within the patterns.

In section~\ref{Method}, we discuss the experimental methods used to collect the data. In section~\ref{Results Pattern Characterization}, we present the observed typical patterns and the corresponding regime diagram. 
After reviewing the Geometric Model's governing equations in section~\ref{GM Review}, we provide a more detailed bifurcation diagram that captures the pattern variations. The model also helps explain the expected phase diagram of the model, which appears in the following section.
In section~\ref{Results Comparison}, we provide a detailed comparison between the new experimental data and the predictions of the GM. We show where the model predictions deviate from the experimental observations. In section~\ref{Disscussion Conclusion}, we summarize our findings and provide suggestions for future avenues of research on this topic.

\section{\label{Method}Experimental Method}
\begin{figure}[htbp]
	\centerline
        {\includegraphics[scale=0.13]{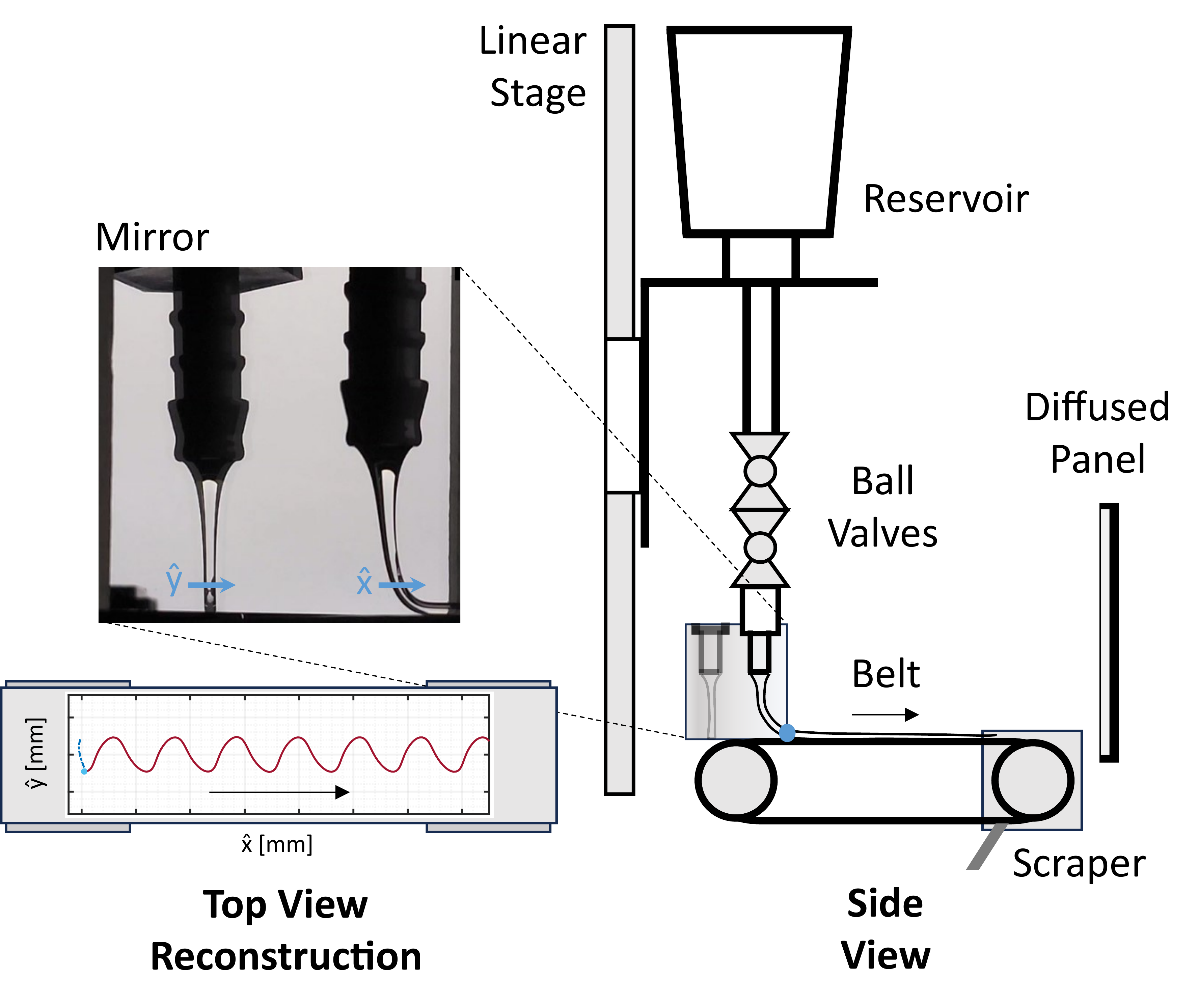}}
	\caption{(Color online) Schematic of the experimental setup. On the top view of the belt, an example of the meander pattern is reconstructed (red, solid line) using the orbit on the left (blue, dashed line). The orbit is obtained by tracking the motion of the thread seen by pointing the camera to the mirror (demonstrated with the $\hat{x}$ and $\hat{y}$ arrows).}
	\label{fig Apparatus}
\end{figure}

The apparatus design drew inspiration from Welch et al.~\cite{welch_frequency_2012}. Fig.~\ref{fig Apparatus} shows the schematic of the setup. We used silicone oil from Dow Corning. The kinematic viscosity $\hat{v}$, density $\hat{\rho}$ and surface tension $\hat{\sigma}$, as provided by the manufacturer's datasheet, are 30~000 cSt, 973 kg/m$^3$ and 21.5 mN/m respectively. The fluid was dispensed from a nozzle with an orifice of 0.805 $\pm$ 0.002 cm. The volumetric flow rate $\hat{Q}$, provided by a gravity feed reservoir, is small enough that the variability in head height is insignificant. Two ball valves were used to control and set the flow. The mass flow rate $\hat{\rho}\hat{Q}$ was set to 0.027 $\pm$ 0.001 g/s and measured before each run using a digital milligram scale. The fall height $\hat{H}$ was controlled using a linear stage precise to $\pm$ 0.05 cm. The fluid was dropped onto a 16 mm-width timing belt acting as a continuous moving surface. The belt was driven directly by a stepper motor attached downstream to provide tension in the upper part of the belt. An encoder was placed upstream to measure $\hat{V}$ to within $\pm$ 0.025 cm/s. The motors ran smoothly with 51~200 steps per revolution provided by TMC2209 motor drivers. The setup was controlled using an Arduino connected to a computer for keyboard inputs. A scraper situated downstream under the belt removed the deposited fluid and maintained a consistent surface for the incoming fluid. A 45-degree mirror, positioned upstream, provided a view of the lateral movements of the thread ($\hat{y}$) while simultaneously observing the longitudinal movements ($\hat{x}$). A lit panel positioned downstream provided diffused lighting for a uniform background behind the fluid thread. The panel was the only source of light, which provided the necessary contrast to the thread. We used a cellphone camera to record 4K videos of the thread movements at 60 fps. It pointed at the mirror capturing the $\hat{x}$ and $\hat{y}$ movement of the thread (see mirror in Fig.~\ref{fig Apparatus}).

Image analysis was performed on a horizontal line above the belt to track the thread motion. The distance from the belt was chosen as small as possible to minimize interference with the deposited fluid. We used a correlation method with a curve fit to interpolate the subpixel thread location.

\begin{figure}[bp]
	\centerline
        {\includegraphics[scale=0.52]{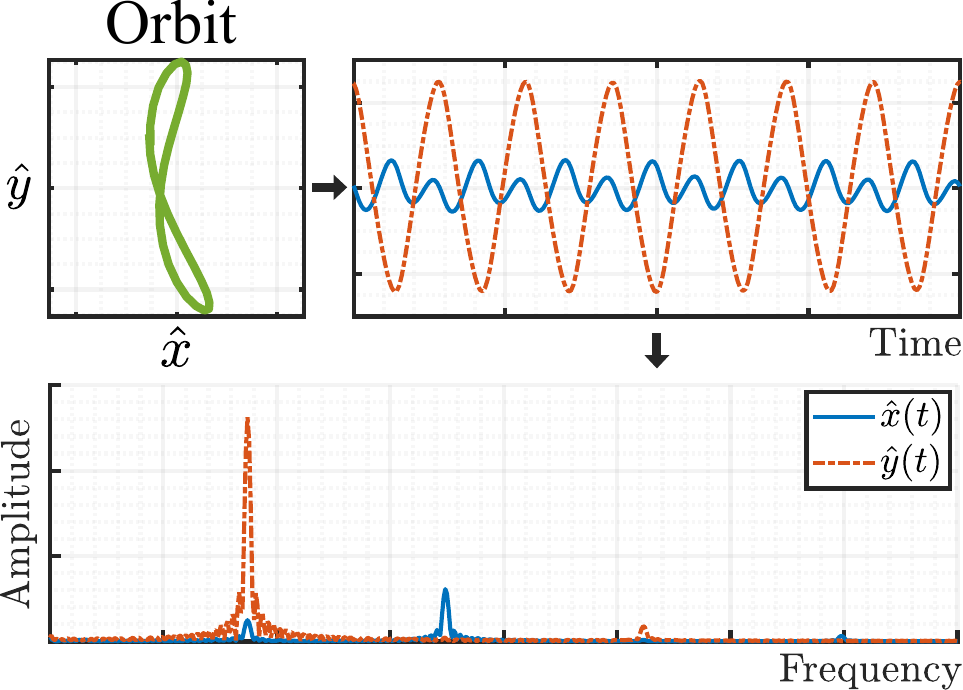}}
	\caption{The meander pattern in the $(\hat{x},\hat{y})$ plane (top left), the time domain (top right), and the frequency domain (bottom). A similar conversion sequence is used for other patterns for classification.} 
	\label{fig FourierMethods}
\end{figure}

Similar to~\cite{welch_frequency_2012} and ~\cite{brun_DVT_2012}, we used FFT on $\hat{x}(t)$ and $\hat{y}(t)$ to characterize and classify the patterns. Each pattern has distinct features in its frequency spectrum, such as relative peak frequencies and amplitudes, making it possible to classify each pattern systematically. An example frequency spectrum is shown in Fig.~\ref{fig FourierMethods} for the meander pattern. Because the criteria for the identification require some initial knowledge of each pattern's frequency spectrum, we added a visual inspection stage to identify atypical characteristics. 

\begin{figure}[tbp]
	\centerline
        {\includegraphics[scale=0.16]{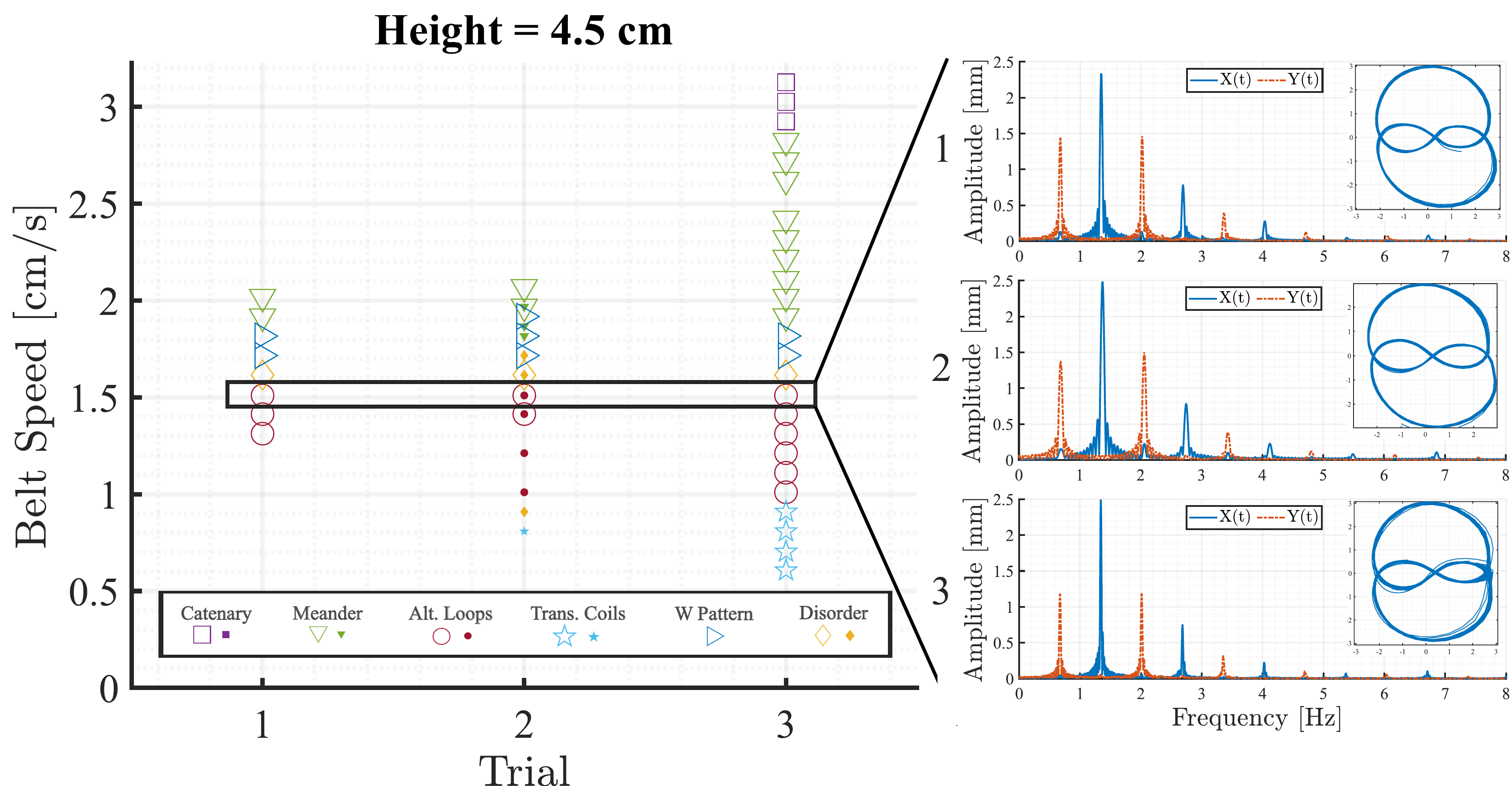}}
	\caption{(Color online) Repeatability of the observed patterns at a fall height of 4.5 cm. Empty markers represent the data collected for an increasing $\hat{V}$ and solid markers are data collected for decreasing $\hat{V}$.}
	\label{fig Val}
\end{figure}

Before collecting data, we obtained preliminary results and benchmarked against the data of \cite{welch_frequency_2012} to validate the experimental methodology. We do so by varying the belt speed at 0.1 cm/s intervals.
The sequence of patterns and the transitions observed in our experiments matched those reported in Welch et al.~\cite{welch_frequency_2012}.
Additionally, we performed multiple measurements to confirm the repeatability of our results. Fig.~\ref{fig Val} shows different tests at $\hat{H}$~=~4.5~cm. The patterns and transitions were consistent throughout the attempts. Furthermore, the patterns taken at the same belt speed near the pattern transition displayed a similar spectrum with peak frequencies and amplitude varying no more than $\pm$ 0.05 Hz and $\pm$ 0.15 mm respectively. 

The experimental data were obtained by varying both the surface speed and the height of the fall around the W pattern observed at the height of 4.5 cm. Both parameter variations were performed in discrete steps of 0.05 - 0.1 cm/s for $\hat{V}$ and at least 0.1 cm for $\hat{H}$. All the data points have a time series ranging from 15 seconds to 110 seconds. 

\section{\label{Results Pattern Characterization}Pattern Characterization}

    \subsection{\label{Results: Basic Patterns}Basic Patterns}
\begin{figure*}[htbp]
        \centerline
        {\includegraphics[scale=0.35]{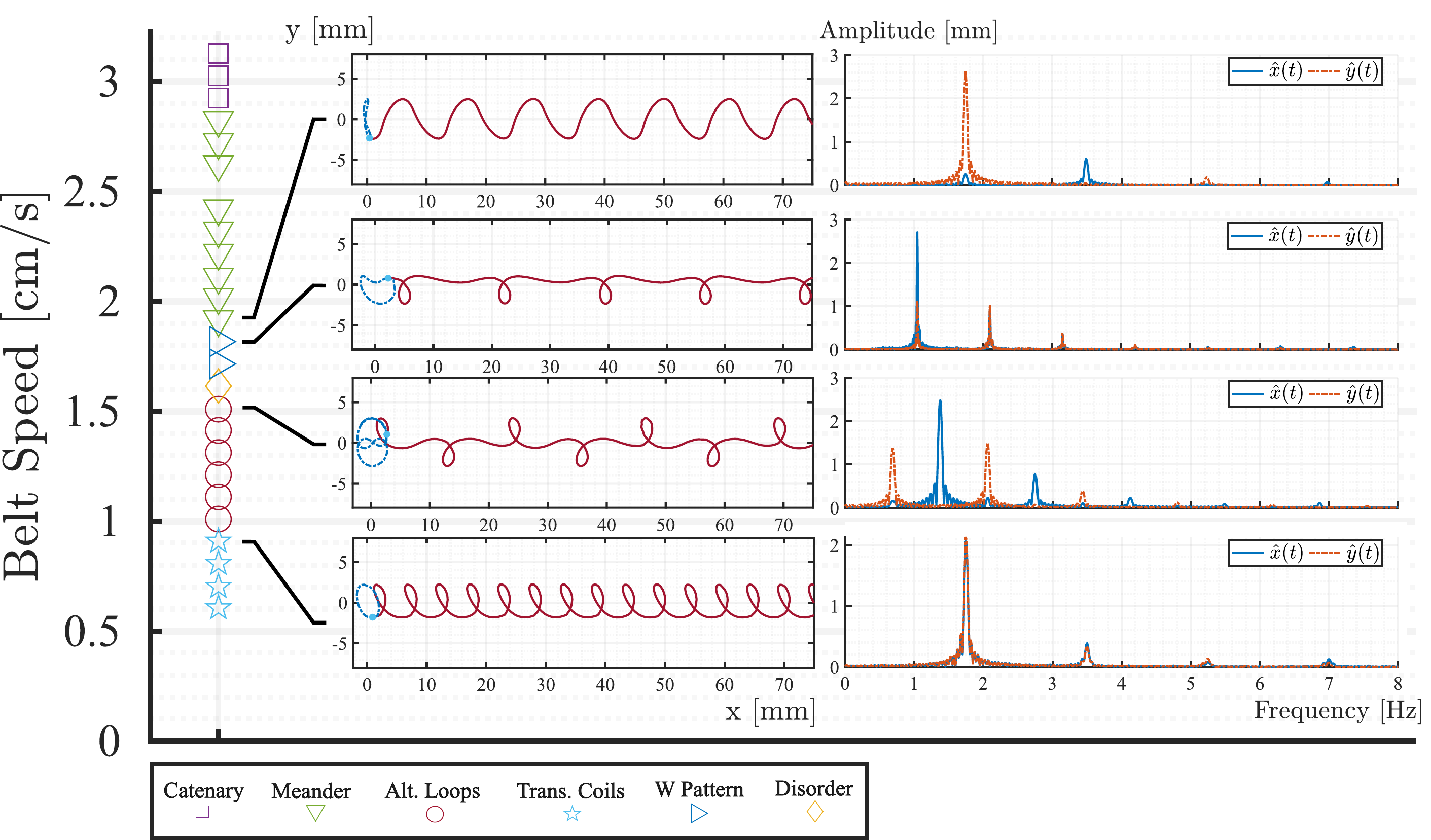}}
        \caption{(Color online) Examples of the basic patterns observed and their frequency spectrum. The orbit and the belt patterns are represented in dashed blue and solid red respectively. The light blue dot represents the thread position at the point this moment in time. Data were obtained at fall height $\hat{H}$ = 4.5 cm for increasing belt speed $\hat{V}$.}
        \label{fig BasicPatterns}
    \end{figure*}

    We review the characteristics of the basic patterns observed from our data. 
    The pattern sequence from the third trial in Fig.~\ref{fig Val} shows the main patterns observed. We take an example of each pattern to show the orbit, belt pattern and frequency spectrum in Fig.~\ref{fig BasicPatterns}.
    The sequence of patterns observed during an increase in belt speed or a decrease in fall height is the following: translating coiling, alternating loops, W pattern, and meander. We include the W pattern as one of the basic patterns. 

    \subsubsection{\label{Results BP: TC}Translating Coiling}
    The translating coiling appears at low belt speeds. The thread moves in the shape of a deformed circle. This is expected as the moving belt breaks the symmetry of a circular orbit from coiling on a stationary surface ($\hat{V} = 0$). 

    The frequency spectrum contains a single prominent peak in both $\hat{x}$ and $\hat{y}$ directions followed by very small peaks at integer multiples from the dominant peak. All peaks in $\hat{x}$ and $\hat{y}$ have similar frequencies and amplitudes.
    
    \subsubsection{\label{Results BP: AL}Alternating Loops} 
    In the alternating loops pattern, the thread travels near the centerline ($\hat{y} = 0$) towards the negative $\hat{x}$ direction. The orbit reveals a single-cycle meandering motion when the thread travels in this direction before initiating a loop. The loops occur on either side and travel away from the centerline towards the positive $\hat{x}$. 

    The frequency spectrum has many well-defined peaks with the largest in the $\hat{x}$ direction and progressively smaller ones at integer multiples of this frequency.
    In the $\hat{y}$ direction, the peaks are smaller than (but of the same order magnitude as) those in the $\hat{x}$ direction, starting at half the frequency of the $\hat{x}$'s largest peak. 
    
    \subsubsection{\label{Results BP: W pattern}W Pattern}
    The W pattern appeared only when increasing $\hat{V}$ or decreasing $\hat{H}$. The pattern resembles a stretched translating coiling but with a meander in between the loops.

    The frequency spectrum is equally as rich as that of alternating loops. The peaks in the $\hat{x}$ and $\hat{y}$ are at the same frequency. The dominant peak is in the $\hat{x}$ direction with progressively smaller peaks at integer multiples away from the dominant frequency. The $\hat{y}$ peaks follow the same trend as the $\hat{x}$, with the exception that the first and second peaks have similar amplitudes.

    We note that the W pattern is not symmetric with respect to the centerline, $y=0$. Coexisting with this pattern is a pattern that is reflected with respect to the centerline and otherwise identical to the W pattern. The coexistence of the two symmetry-broken patterns is because of the invariance of the system with respect to the centerline (if one pattern exists then the other one should too). In the rotational version of the experiments~\cite{lisicki_viscous_2022}, for example, the centripetal acceleration is expected to break this invariance. In our experiments, we observed both of these patterns, although one of them appeared more frequently. We take the appearance of the two symmetry-broken patterns as an indication that the moving belt did not lean significantly toward one direction.

    \subsubsection{\label{Results BP: Meander}Meanders} 
    The meandering pattern occurs at larger belt speeds. The orbit resembles a flattened figure-eight. As the meander approaches the straight pattern (increasing belt speed), the thread motion becomes almost entirely in the $\hat{y}$ direction with decreasing amplitudes. The pattern shape is not purely harmonic (slightly tooth-like) possibly due to movements in the thread not reflecting the traced path on the belt. 

    The frequency spectrum consists of a dominant frequency in the $\hat{y}$ direction and two smaller frequencies in the $\hat{x}$ direction: one matching the dominant frequency and the other (largest) at twice the dominant frequency; recall the time-domain trace in Fig.~\ref{fig FourierMethods}. 
    
    \subsection{\label{Results Regime Diagram}Regime Diagram}
\begin{figure*}[htbp]
	\centerline
        {\includegraphics[scale=0.60]{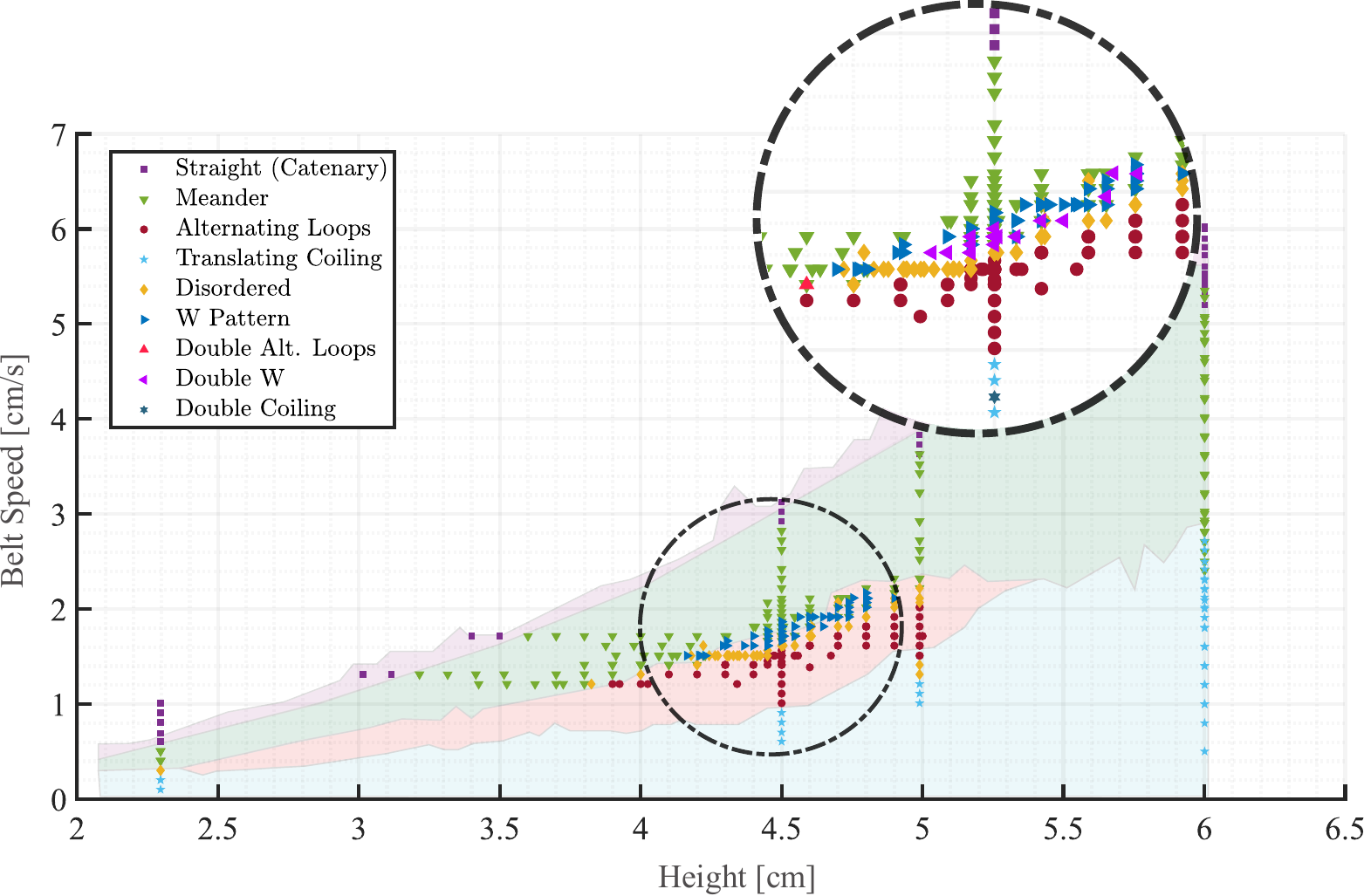}}
	\caption{Dimensional regime diagram   with the patterns collected in terms of $\hat{V}$ and $\hat{H}$. Inset enlarges the W pattern region with period doubling patterns made visible. In the background is the simplified regime diagram from Welch et al. \cite{welch_frequency_2012}.}
	\label{fig DimensionalRD}
    \end{figure*}

    We consolidate all the collected data into a regime diagram as shown in Fig.~\ref{fig DimensionalRD}. We included a simplified version of the diagram of \cite{welch_frequency_2012} in the background for comparison. We observed similar results to those reported in \cite{welch_frequency_2012}. 
    
    The alternating loops region occupies a limited area between meandering (at higher $\hat{V}$) and translating coiling (at lower $\hat{V}$). The region disappears at low and high values of $\hat{H}$, making way for a direct transition between meander and translating coiling. The transition between meandering and translating coiling exhibits hysteresis. Additionally, the region corresponding to the W pattern region, also called the \textit{one side-looping state}~\cite{welch_frequency_2012}, was locally sandwiched between the meandering and the alternating loop patterns around $\hat{H}=4.5$ cm. We note that the W pattern region occupies a small yet significant space in the diagram; previous regime diagrams did not include this region~\cite{welch_frequency_2012}.

    Compared to the other three patterns, the W region was small and only appeared when approached from the alternating loops region. We found the W region to be comprised of two distinguishable patterns; see the inset of Fig.~\ref{fig DimensionalRD}. These patterns are the W pattern and a period-doubled W pattern, which we call double W. To our knowledge, the double W has never been reported before. This pattern appears to coexist with the W pattern around the same parameter space but in a smaller region.
    We also note appearances of double alternating loops\footnote{This pattern was named \textit{double-8} by Chiu-Webster and Lister \cite{chiu-webster_fall_2006} because it is the period-doubled version of the \textit{figure-of-8} pattern. The figure-of-8 pattern was later renamed to alternating loops~\cite{brun_DVT_2012,welch_frequency_2012}, which is the name we adopt in this work.} and double coiling which is surprising because they were only previously observed at higher heights \cite{chiu-webster_fall_2006}. We discuss these period-doubled patterns in section~\ref{Results: PD Patterns}.
    Moreover, we frequently observed disordered patterns along the boundaries of the alternating loop pattern; these are discussed in section~\ref{Results: Disordered}.

\section{\label{GM Review}Review of the Geometric Model}
In this section, we revisit the geometric model of coiling (GM) and provide a bifurcation diagram that elucidates certain aspects of the transitions between the patterns. The details of the derivation can be found elsewhere~\cite{brun_GM_2015}. 

\subsection{Geometric Model of Coiling}

\begin{figure}[htbp]
    \centerline
    {\includegraphics[scale=0.1]{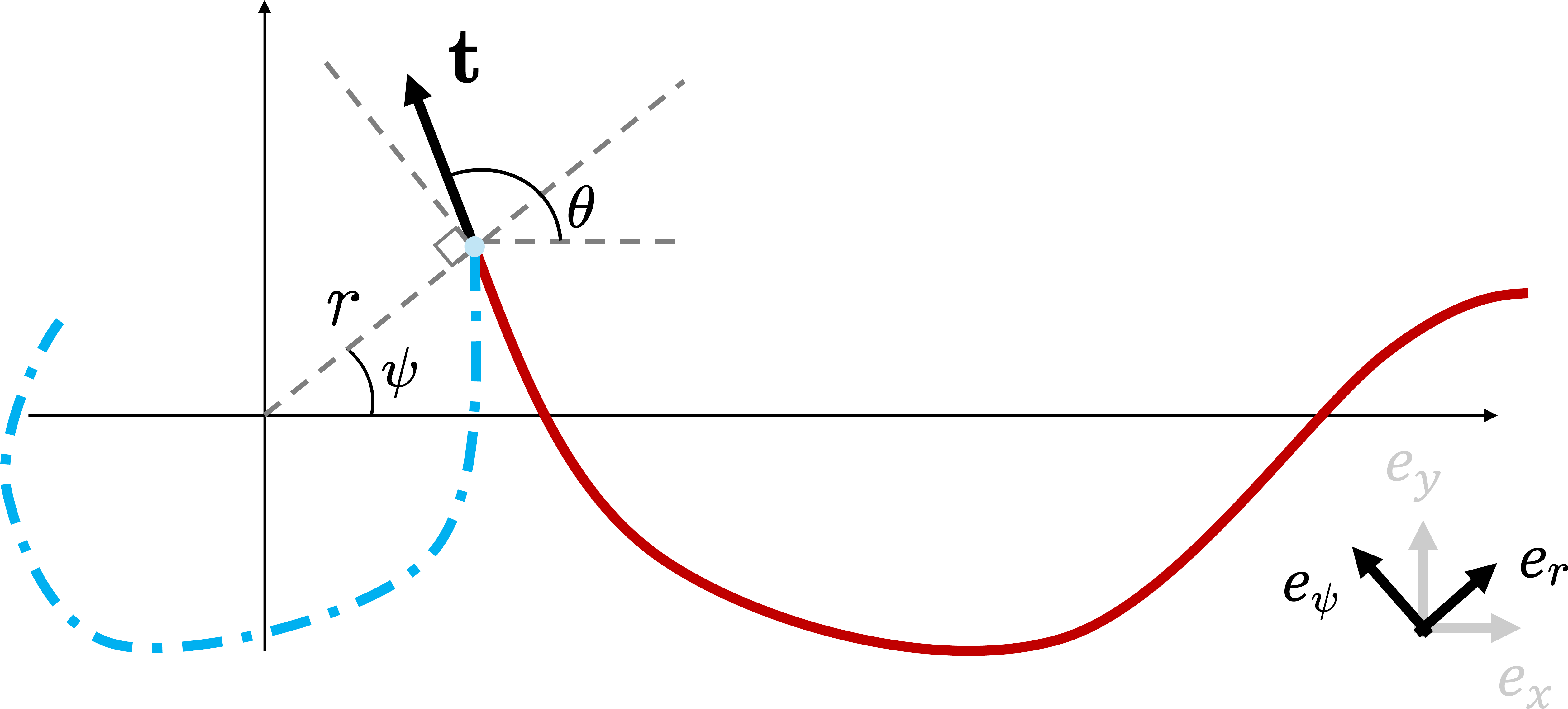}}
    \caption{Schematic of the orbit (dashed blue) and the belt pattern (solid red) with parameters $r$, $\psi$, $\theta$ and $\textbf{t}$ describing the thread motion at the contact point. $\textbf{t}$ is the tangent of the belt pattern at the point of contact. Source: Adapted from Fig.~3 of \cite{brun_GM_2015}}
    \label{fig GMSketch}
\end{figure}

The GM is a simplified model based on the geometrical and local bending properties of the moving heel. It is derived by enforcing a no-slip boundary condition between the deposited fluid and the moving belt. This results in a system of three ODEs that can successfully describe the basic four patterns at small heights, where inertial effects are negligible. The model can be made dimensionless using parameters corresponding to coiling on a stationary surface. In this specific coiling regime, the thread moves in a circular orbit with tangential speed defined as $\hat{U}_C = \hat{R}_c\hat{\Omega}_c$, where $\hat{R}_c$ is the radius of the coil and $\hat{\Omega}_c$ is the coiling frequency. 
Using these parameters to scale the equations, we have
\begin{subequations}
    \begin{align}
    r^\prime & = \cos(\theta-\psi) + V\cos(\psi) \\ 
    r\psi^\prime & = \sin(\theta-\psi) - V\sin(\psi) \\ 
    \theta^\prime & = \kappa(r,\theta-\psi)
    \end{align}
    \label{eqn: Geometric Model ODE}
\end{subequations}
where ($r$,$\psi$) are the polar coordinates of the contact point from the projection of the nozzle on the belt at its origin, $V$ is the ratio between $\hat{V}$ and $\hat{U}_c$, and $\theta$ is the direction of the tangent $\textbf{t}$ to the heel described by its curvature $\kappa$ (see Fig.~\ref{fig GMSketch}). Because of conservation of mass, $\hat{U}_c$ is equivalent to the free-fall speed of the liquid moments before contact. Thus, the region where coiling patterns emerge correspond to $V\le1$. The prime denotes differentiation with respect to the arclength of the thread. 
Parameters $r$ and $\kappa$ are made dimensionless by dividing or multiplying them by $R_c$. 
The equation for $\kappa$ is obtained from a fit to the data from simulations using the DVT as
\begin{equation}
    \kappa(r,\theta-\psi) = \sqrt{r}(1+A(\theta-\psi)r)\sin(\theta-\psi)
    \label{eqn: Heel Curvature}
\end{equation}
Here, $A(\theta-\psi) = b^2\cos(\theta-\psi)/(1-b \cos(\theta-\psi))$ where $b=0.715$. See the supplementary material in~\cite{brun_GM_2015} for further details on obtaining the curvature equation. 
    
Note that the GM, Eq. (\ref{eqn: Geometric Model ODE}), is governed by a single parameter, $V$. The non-inertial patterns are thus only dependent on $V$ and not on the fall height. We will revise the regime diagram based on this information in section~\ref{Results Comparison}.

\subsection{Bifurcation Analysis}

At a given value of $V$, the GM is integrated in time until a steady-state periodic solution is obtained. This is the coiling pattern in the $(r,\psi)$ coordinate system. For some values of $V$, multiple stable patterns coexist and the ensuing pattern depends on the initial conditions. A bifurcation diagram helps visualize the sequence of, and transitions between, the patterns, especially in these situations. Brun et al.~\cite{brun_GM_2015} presented a simplified bifurcation diagram that showed the ranges of $V$ corresponding to the four basic patterns. Here, we present a more detailed bifurcation diagram to better understand the morphology of the patterns. We use {\sc auto-07p} to perform a bifurcation analysis of the patterns predicted by the GM as families of periodic orbits that satisfy a suitable two-point boundary-value problem~\cite{auto,Sebius_RedBook}. 

\begin{figure}[tbp]
        {\includegraphics[scale=0.195]{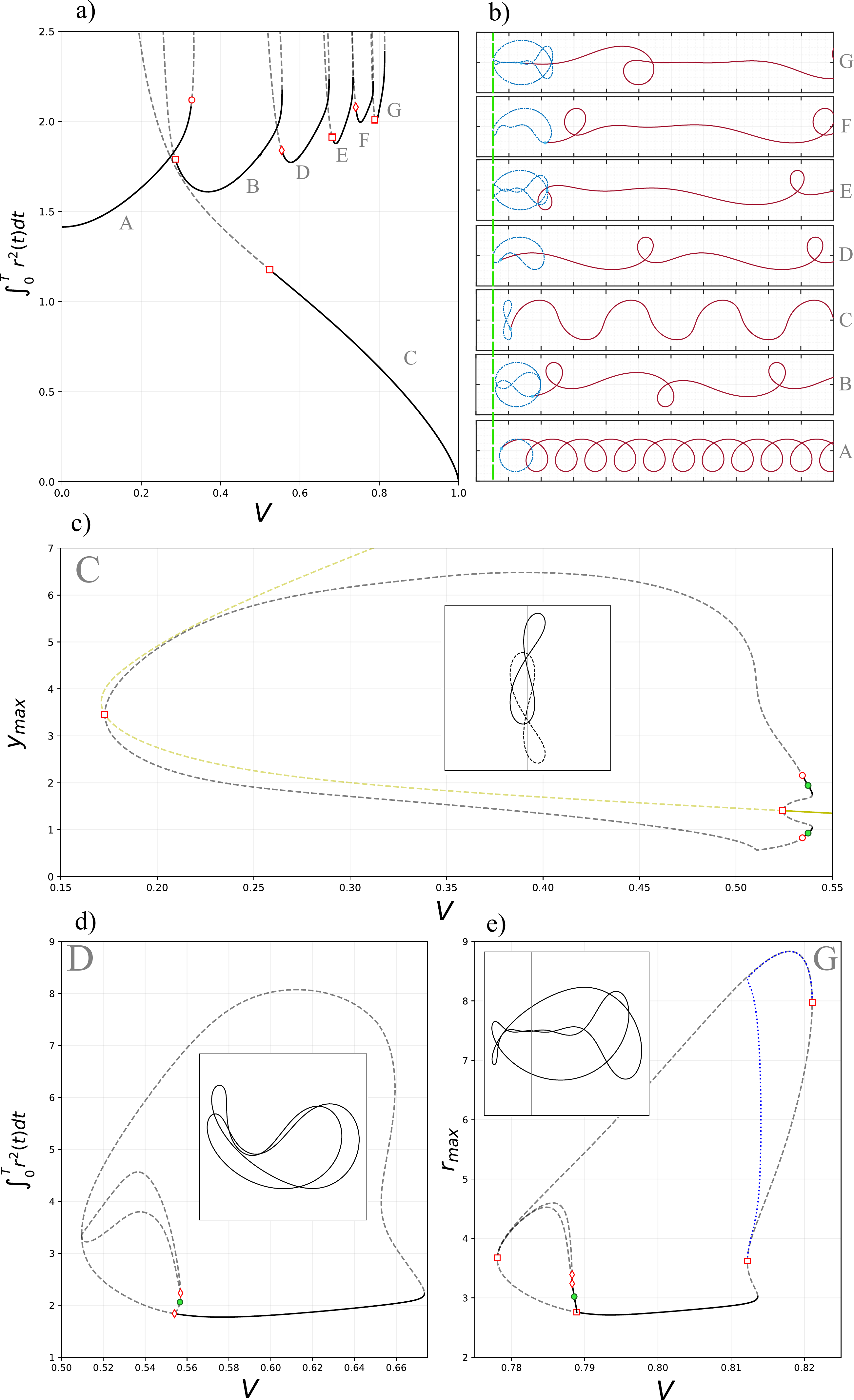} \hfill 
        }
    \caption{a) Bifurcation diagram and b) the corresponding patterns from the Geometrical Model. The patterns are A: translated coiling, B: alternating loops, C: meandering, D: W pattern, E and G: variations of alternating loops, F: variation of W pattern. The green dashed line ($r_{crit}$) is the distance where looping is believed to initiate. Diamond, square and circle markers indicate period-doubling, branch point and torus bifurcations, respectively. Unstable solutions are indicated using gray dashed lines. c), d) and e) give detailed views of branches C, D and G respectively. In c), the yellow branch is the main branch appearing in a). Insets show the orbit at the green data points.}
    \label{fig Bifurcation}
\end{figure}

Fig.~\ref{fig Bifurcation}(a) shows the bifurcation diagram in the $(V,r_{N})$ plane, with $r_{N}=\int_0^Tr^2(t)dt$ where $T$ is the period of the pattern. 
The diagram indicates several solution branches, labelled A to G, each corresponding to a different pattern in Fig.~\ref{fig Bifurcation}(b). We only report the solution branches with stable patterns. The stability of each pattern is determined by the Floquet multipliers of the corresponding periodic orbit. The solid portions of the curves correspond to (linearly) stable patterns and the dashed portions indicate unstable patterns. 

The translated coiling pattern, branch A, starts from the static coiling pattern at (0, $\sqrt{2}$). The pattern loses stability via a torus bifurcation near $V\approx0.325$. The remainder of the branch, as well as the pattern emanating from the torus bifurcation, remain unstable. 

The stable portion of the alternating loops pattern, branch B, starts near $V\approx0.289$ at a branch point (symmetry-breaking bifurcation) and terminates near $V\approx0.555$ through a turning point (saddle-node bifurcation). Thus, the alternating loops pattern overlaps with the translated coiling pattern at its lower velocities and with the meander pattern (branch C) at its higher velocities. The stable portion of the meander pattern, starting at a branch point near $V\approx0.524$, extends to (1, 0) where it terminates into the straight (catenary) pattern.

Fig.~\ref{fig Bifurcation}(c) shows the two branches (black curves) that emerge from the symmetry-breaking bifurcation point of the meander pattern on branch C (main branch shown in yellow). To distinguish the two coexisting branches of patterns from each other, we have used $y_{max}$ on the vertical axis, which represents the largest distance of each pattern from the $x$ axis. There is a very small portion of these branches that is stable, occurring between a turning point and a torus bifurcation. The inset shows the patterns on each branch at a point along this stable portion near $V\approx 0.538$. As expected, reflecting one of the symmetry-broken patterns with respect to the $x$ axis produces the other pattern. 

The W pattern, branch D, appears in the immediate vicinity of the alternating loops pattern. Its stable portion starts at a period-doubling bifurcation near $V\approx0.554$, very close to the turning point of branch B, and terminates at a turning point near $V\approx0.673$. The W pattern overlaps with the meander pattern over its entire range of belt velocity, a feature that makes its experimental observation more challenging than the meander pattern. A period-doubled version of the W pattern is repeatably observed in experiments near the period-doubling bifurcation point, which will be discussed in Sec.~\ref{Results: PD Patterns}.

Fig.~\ref{fig Bifurcation}(d) shows a more detailed closeup of the solution branch for the W pattern. Specifically, a branch of period-doubled patterns emanates from the period-doubling bifurcation point; the inset shows the pattern for one point along the branch. This branch of period-doubled patterns undergoes another period-doubling bifurcation (shown in blue). It is conceivable that these two branches indicate the beginning of a period-doubling route to chaos. We did not investigate this in detail because the ensuing patterns remain unstable.

Several other patterns exist beyond the W pattern that resemble stretched versions of the alternating loops and W patterns. The bifurcation diagram in Fig.~\ref{fig Bifurcation}(a) shows only three of these patterns, branches E to G, all of which include stable portions. These patterns have not been reported in experiments or DVT simulations yet, though similar stretched patterns were captured in simulations with the elastic equivalent of the DVT~\cite{jawed_geometric_2015}. 

All of the branches A to G extend to much higher values of $r_{N}$, undergoing various bifurcations. These patterns remain predominantly unstable and are not discussed here in detail. As an indicative example, Fig.~\ref{fig Bifurcation}(e) shows the bifurcation diagram for branch G. The primary stable portion of the pattern extends from a symmetry-breaking bifurcation near $V\approx0.789$ and extends to near $V\approx0.814$ where it loses stability through a turning point. Interestingly, there is a small portion of the branch emanating from $V\approx0.789$ that is stable; the inset shows the corresponding orbit at a point along the branch. This pattern loses stability through a period-doubling bifurcation. Similar to the situation for the W pattern, this may indicate a period-doubling route to chaos in branch G. We did not investigate this in detail because the ensuing patterns are unstable.

The pattern sequence in the upper branches (A, B, D, E, F, G) of Fig.~\ref{fig Bifurcation} appears to be related to an expanding meandering motion of the thread contact point when travelling against the belt direction (upstream). For every increase in belt speed at the same height, the contact point travelling upstream becomes slower because the fluid speed remains the same. With a slower longitudinal travel, the thread will oscillate longer in the transverse direction. Moreover, there seems to be an approximate common distance $r_{crit}$ in the negative $x$ where a loop initiates with the thread returning in the positive $x$ direction (downstream); see Fig.~\ref{fig Bifurcation}(b). This may be related to the heel requiring a certain curvature to perform a loop. The heel needing to reach this critical distance before initiating a loop also contributes to the added oscillation of the meander. For an increasing belt speed, the system will transition from one pattern to another after an extra half oscillation is added to the upstream meandering motion resulting in patterns of type alternating loops (B, E and G) and W pattern (D and F).
    
Finally, we note that the various patterns (solution branches) are disconnected from each other in the bifurcation diagram; {\it i.e.} they appear as {\it isolas}. We have not explored the connection between these patterns. The topology of the orbits, and the existence of periodic orbits with very long periods, strongly hints at the existence of global bifurcations that organize (at least some of) the orbits. This analysis falls beyond the scope of the current paper and will be reported elsewhere.

\section{\label{Results Comparison}Comparison with GM} 
To compare with the GM, we study the dimensionless form of the regime diagram, the frequency and amplitude variations of the Fourier spectrum peaks, and the time period of the patterns. 
All results have been rescaled by dividing with either $\hat{U}_C$, $\hat{R}_c$ or $\hat{\Omega}_c$, as suitable. The sole exception is the dimensionless height $H$, for which we divide $\hat{H}$ by a length of scale $(\hat{\nu}^2/\hat{g})^\frac{1}{3}$ following \cite{brun_DVT_2012,brun_GM_2015}.
We used $\hat{V}_{crit}$ and the corresponding meandering frequency for $\hat{U}_C$ and $\hat{\Omega}_c$; these are equivalent as already shown in \cite{ribe_stability_2006,brun_GM_2015}.

    \subsection{\label{Results: Dimensionless RD}Dimensionless Regime Diagram}
\begin{figure}[htbp]
	\centerline
        {\includegraphics[scale=0.38]{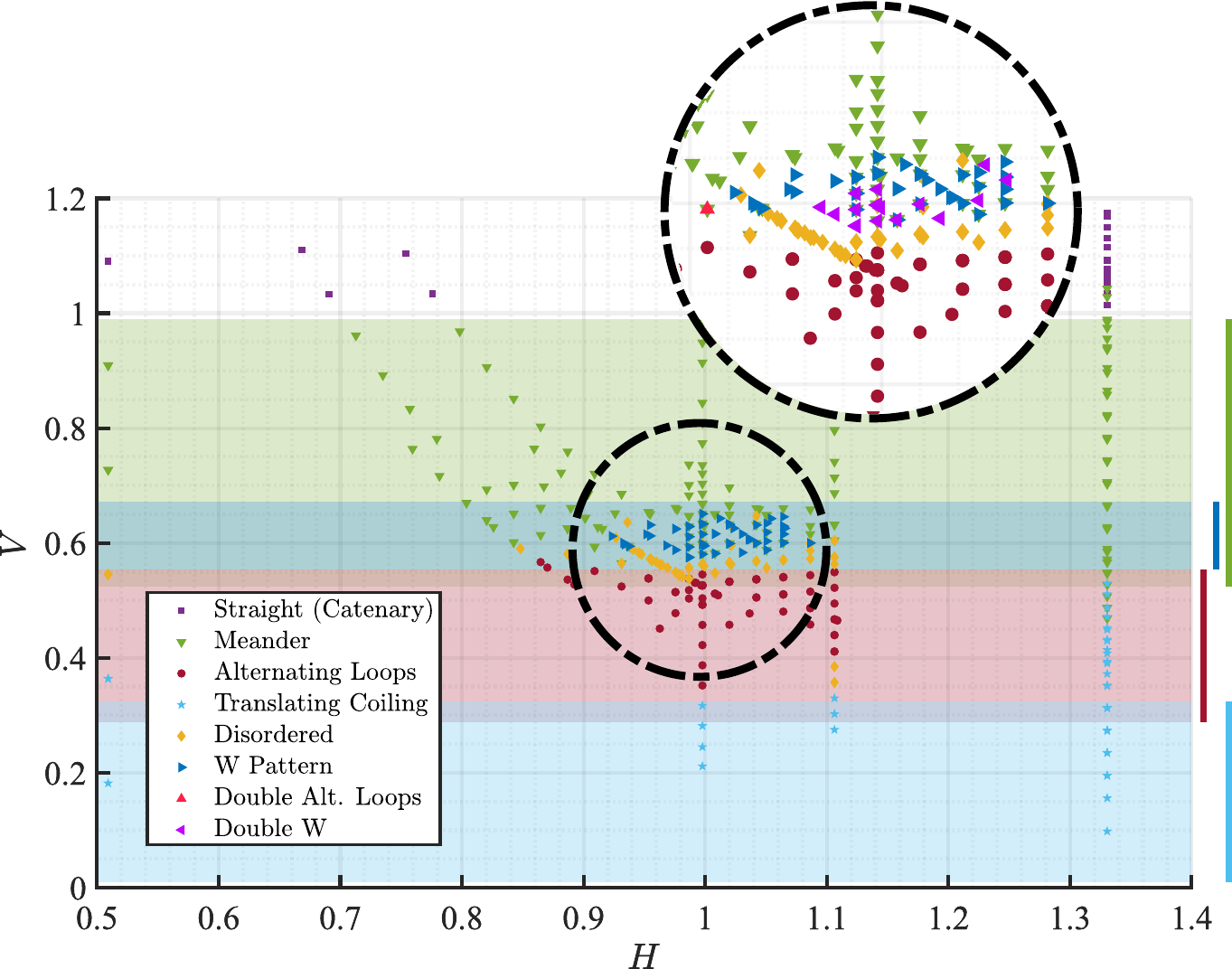}}
	\caption{A comparison of the experimental data (opaque), scaled into its dimensionless form, with the predictions of the GM (faded) for an increasing $V$. The regions are coloured to match the markers of the corresponding patterns in the legend. Inset enlarges the W pattern region with period doubling patterns made visible. The coloured vertical lines on the right indicate the extent of each of the predicted patterns over $V$. $V = \hat{V}/\hat{U}_c$ and $H = \hat{H}(\hat{g}/\hat{\nu}^2)^\frac{1}{3}$}
	\label{fig GM Comparison}
    \end{figure}

    The experimental data from Fig.~\ref{fig DimensionalRD} are rescaled into dimensionless parameters using $V$ and $H$. Fig.~\ref{fig GM Comparison} shows the dimensionless regime diagram. The coloured horizontal regions show the predictions of the GM.
    For most of the data, the GM correctly identifies the regions where the patterns appear in the experiments. The borders between translating coiling, alternating loops and meandering appear to follow the transition points predicted by the GM.
    The region where the W pattern occurs is observed primarily near $H\approx 1$ and does not extend across the full range of $V$, both in contrast to the predictions made by the GM. The range in $V$ for the W pattern becomes narrower at lower and higher $H$ until eventually, the disordered state takes place.  At the very extremities in the range of $H$ shown, the alternating loop region also becomes narrower (as indicated by the results of Welch et al.~\cite{welch_frequency_2012} in Fig.~\ref{fig DimensionalRD}) until it disappears, leaving a direct transition between meander and translating coiling. 
    Brun et al.~\cite{brun_DVT_2012} showed that above $H \approx 1.3$, the system enters into the inertia-gravitational regime, deviating from the assumptions of the GM. At lower heights, however, inertia is expected to have negligible effects; yet the GM does not correctly explain the observed behaviour of the patterns.

    The period-doubled and disordered patterns are discussed next.
    
        \subsubsection{\label{Results: PD Patterns}Period-Doubled Patterns}
\begin{figure}[tbp]
        	\centerline
                {\includegraphics[scale=0.26]{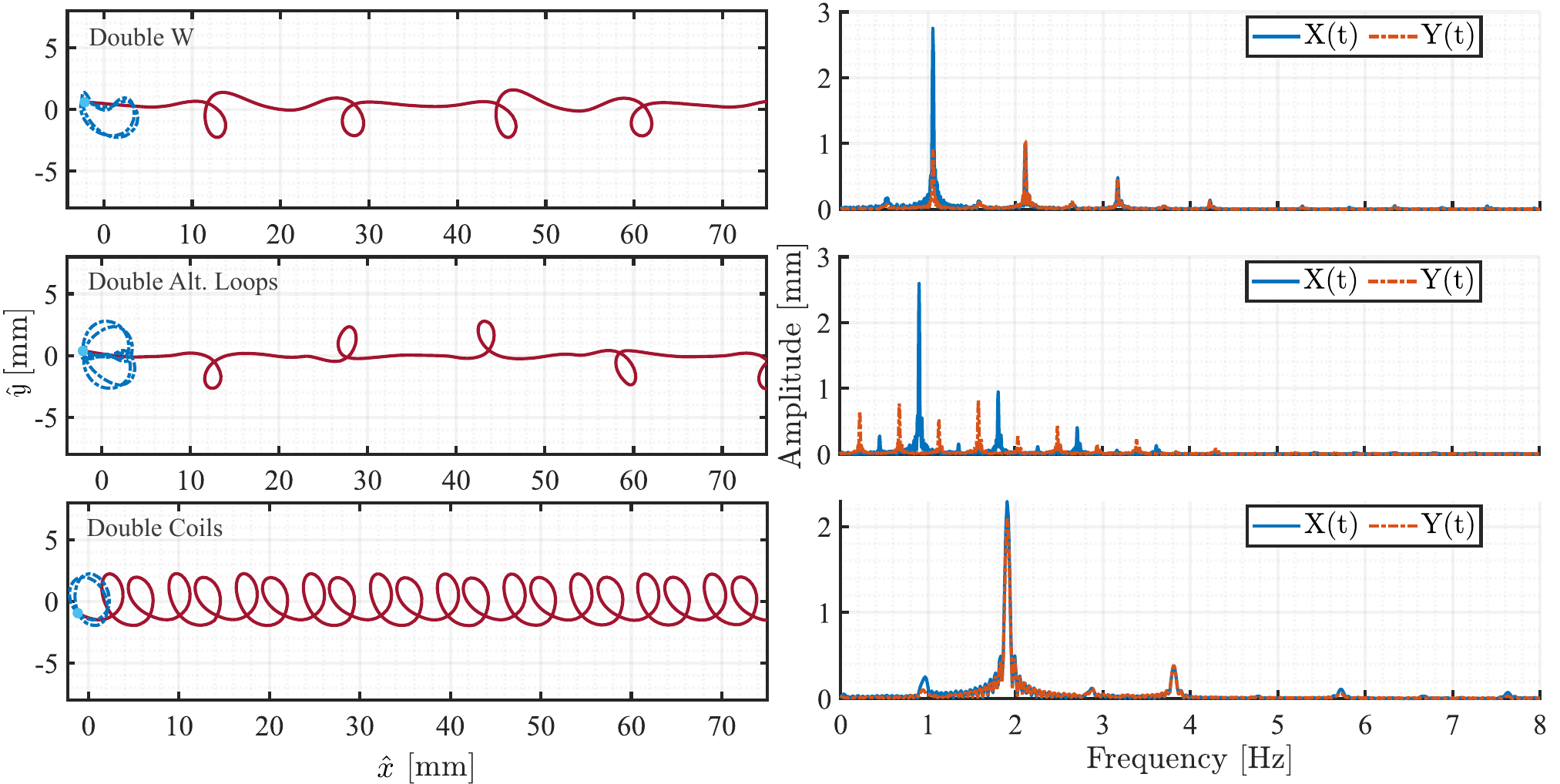}}
        	\caption{(Color online) Period doubled patterns observed from experimental data (right) with their frequency spectrum (left). The orbit and the pattern are represented in dashed blue and solid red respectively. }
            \label{fig PeriodDoubled}
        \end{figure}

        We observed three period-doubling patterns: double W, double alternating loops and double coiling. Because these patterns are intuitively associated with the dynamics of falling liquid rope, they are also called resonant patterns~\cite{brun_DVT_2012, brun_GM_2015}.  Fig.~\ref{fig PeriodDoubled} shows the period-doubled patterns and their frequency spectra. 

        The double W pattern appears to be a period-doubled version of the W pattern. It occurred frequently and repeatably in our experiments, as discussed in section~\ref{Results Pattern Characterization}. The double W pattern is generally obtained at a smaller value of $V$ than the W pattern. The pattern 
        resembles a mixture of the stretched translating coiling and the W pattern seen in DVT \cite{brun_DVT_2012}. Its frequency spectrum is similar to the regular W pattern but with additional smaller peaks between the dominant peaks.

        The W pattern undergoes a period-doubling bifurcation near $V\approx0.555$, as indicated in Fig.~\ref{fig Bifurcation}(c) and reported in~\cite{brun_GM_2015}. While this bifurcation occurs in the proximity of where the double W pattern is borne, comparison of the measured orbit (Fig.~\ref{fig PeriodDoubled}, top row) and the period-doubled orbit predicted by the GM (Fig.~\ref{fig Bifurcation}(c), inset) indicate that the two patterns are not the same. More importantly, the period-doubled pattern predicted by the GM is unstable and cannot be recreated experimentally. We were not able to find any indication of a stable double W pattern in the GM. 

        The double alternating loops and double coiling are resonant forms of the alternating loops and translating coiling respectively. The double alternating loops form two consecutive loops on one side before transitioning to the opposite side.
        The loops are further apart than previously observed by Chiu-Webster and Lister \cite{chiu-webster_fall_2006}.
        This difference is possibly due to the different heights at which the pattern was observed: we observed this pattern at a lower height. The lower fall heights are generally expected to bring about slower dynamics (no inertial effects) in contrast to the faster and more complex dynamics occurring at higher heights. 
        Compared to the alternating loops, the main frequency peaks in the $y$ direction are split into two smaller peaks, leading to the two successive loops. 
    
        The double coiling alternates between two different loops. The overall pattern shape is qualitatively similar to the findings of the DVT \cite{brun_DVT_2012}, which are visibly different from the experimental observations of \cite{chiu-webster_fall_2006}. The frequency contents of the double coiling pattern are similar to those of the translating coiling but, as with the double W, smaller peaks are present between the dominant peaks.

        Double alternating loops and double coils were previously observed at larger heights where inertia starts to become dominant~\cite{chiu-webster_fall_2006, morris_meandering_2008}. Yet based on our observations, it seems they can also occur at lower heights, albeit rarely as they were only observed once during a decreasing $\hat{H}$.  
        The resonant patterns appearing at higher heights were shown previously to coincide with the ranges of height where $\bar{IG}$ is present in LRC \cite{chiu-webster_fall_2006, brun_DVT_2012}. As such, they were thought to be the result of the interaction between the oscillations of the heel with the oscillations of the upper part known as the ``tail''~\cite{chiu-webster_fall_2006,brun_DVT_2012}.    
        Having observed these patterns at lower heights, it may be that inertia, at least in some special cases, can still play a significant role. A deeper look into the dynamics of the tail in experimental data could provide better insight. 
    
        \subsubsection{\label{Results: Disordered}Disordered Patterns}
        In previous studies, the disordered patterns were observed at higher heights and occupied a large region in the regime diagram. Within the regions previously identified as disordered, the pattern appears random and chaotic with a rich Fourier spectrum \cite{morris_meandering_2008, brun_DVT_2012}.
        The term disordered has been defined very loosely in the literature. Chiu-Webster and Lister \cite{chiu-webster_fall_2006} first identified them as \textit{irregular} patterns which consisted of assortments of the periodic patterns observed. In later studies, the name was changed to disordered \cite{morris_meandering_2008,welch_frequency_2012,brun_DVT_2012}. 
        In our case, we classified patterns as disordered if they did not match the characteristics of basic patterns throughout the entire observation period. 

\begin{figure*}[tbp]
    	\centerline
            {\includegraphics[scale=0.47]{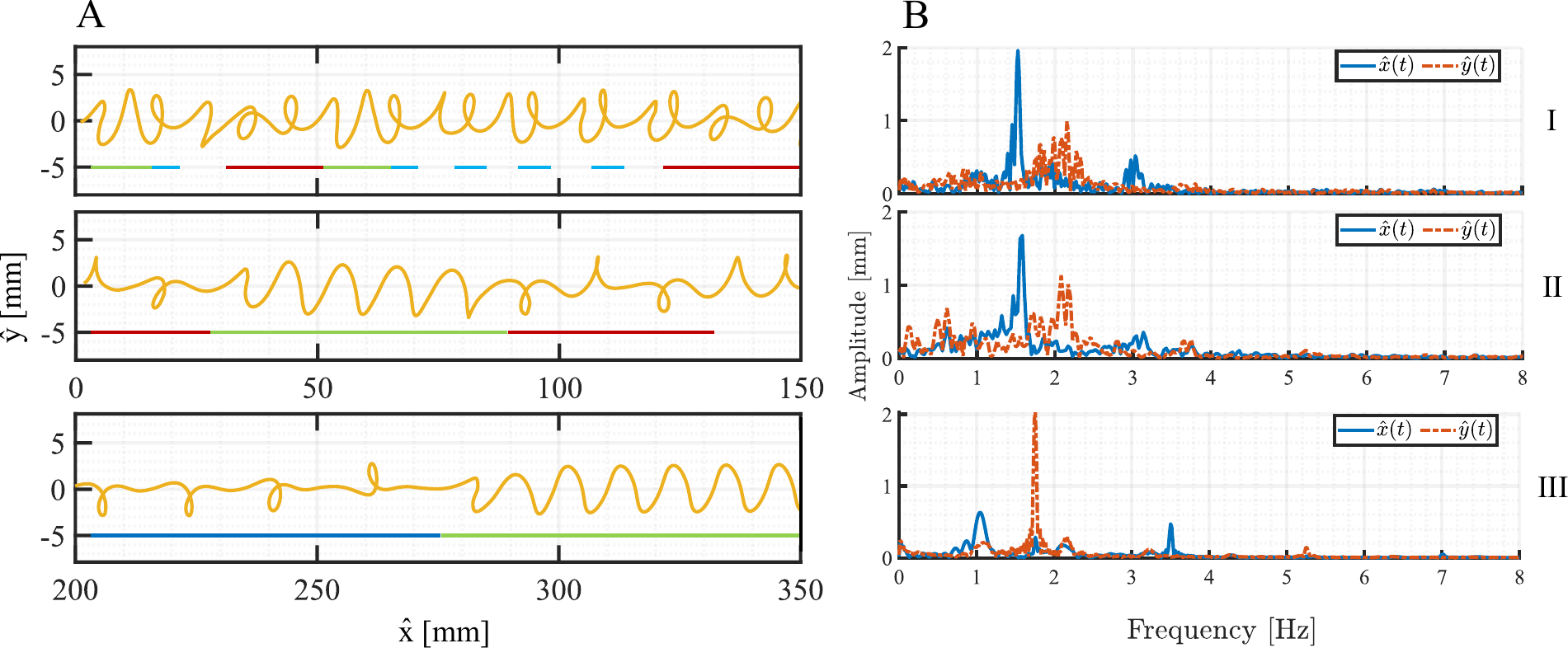}}
    	\caption{Three examples of patterns classified as disordered. The patterns in rows I, II and III are found between regions of translating coiling and alternating loops;  between alternating loops and meanders; and between W state and meander respectively. A) shows the belt patterns of the three types of disordered with sections of basic patterns indicated by lines green (meander), dark blue (W pattern), red (alternating loops), and light blue (translating coiling). B) is the FFT corresponding to the patterns in A).
        }
    	\label{fig Disordered}
        \end{figure*}

        We observed disordered patterns at the boundaries of the alternating loops and W pattern. This transition point coexists with the meander pattern; recall Fig. 8(a). We highlight three types of disordered patterns shown in Fig.~\ref{fig Disordered}. The pattern in panel I of Fig.~\ref{fig Disordered} was seen between translating coiling and alternating loops. The thread appeared to transition rapidly between translating coiling, alternating loops and meandering. 
        The pattern in panel II can be found when transitioning from alternating loops to meanders. The two patterns alternate, each undergoing a few cycles before transitioning to the other. Panel III shows the W pattern becoming a meander after many stable oscillations.

        The appearance of irregular patterns of the types described at the boundaries of alternating loops as shown in Fig.~\ref{fig DimensionalRD}  suggests a higher sensitivity to disturbances in this area.
        Considering that type II and III disordered patterns arise at the borders of the basic patterns they are composed of, it may be more appropriate to name them {\it mixed patterns} instead. 
        These mixed patterns resemble chaotic trajectories that travel between different basins of attraction before latching on to an attractor (a limit cycle, in this case). This did not occur in experiments but was always the case in our numerical studies of the GM. It is not possible at this point to irrefutably conclude the fate of the mixed patterns of type II and III.  
        
    \subsection{\label{Results: Frequency}Frequency}
\begin{figure*}[htbp]
	\centerline
        {\includegraphics[scale=0.65]{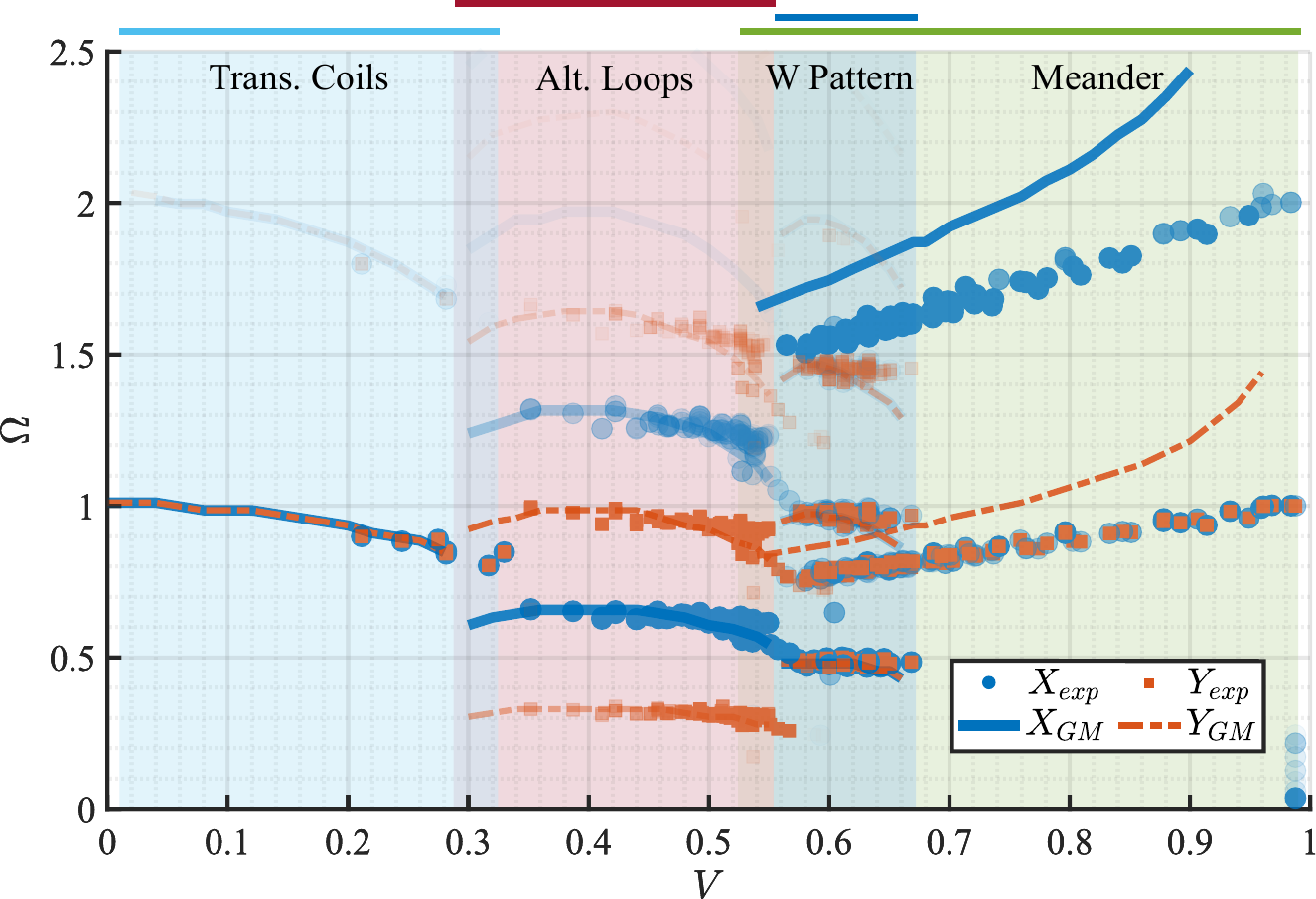}}
	\caption{(Color online) Dimensionless peak frequencies from the Fourier spectrums of all collected patterns around the W region where $\Omega = \hat{\Omega}/\hat{\Omega}_c$. The transparency of the data indicates the relative amplitudes of the peaks. Opaque markers indicate large peaks. The coloured horizontal lines above represent the extent of the predicted pattern regions over $V$.}
	\label{fig Freq}
    \end{figure*}

    The frequencies of the main peaks $\hat{\Omega}$ from the Fourier spectrum of the basic patterns are collected for comparison with the GM. Fig.~\ref{fig Freq} shows the non-dimensional frequency, $\Omega = \hat{\Omega}/\hat{\Omega}_c$, as a function of $V$. The degree of transparency of the data points corresponds to the relative size of their peaks; {\it e.g.} the translating coil exhibits a dominant frequency, with a much weaker peak occurring at the second harmonic. The predictions of the GM are displayed as solid and dashed curves, respectively corresponding to the motion along the $x$ and $y$ axes. The coloured regions represent the patterns predicted by the GM. 
    
    First focusing on the experimental results, we observe that the frequency contents of the patterns exhibit a single dominant peak along both $x$ and $y$ axes in the translating coiling region. In the alternating loops region, the frequency contents split across various peaks approximately at $n\hat{\Omega}_c/3$ where $n \in \{1, 2, 3, ...\}$; also noted by Brun et al.~\cite{brun_DVT_2012}. The W pattern exhibits its dominant peaks at $n\hat{\Omega}_c/2$. The peaks start near $3n\hat{\Omega}_c/4$ for the meander pattern, and increase approximately linearly to $n\hat{\Omega}_c$ until reaching the straight pattern at $V = 1$. The linear variation within the meander pattern is consistent with the experimental observations of Morris et al.~\cite{morris_meandering_2008} and Welch et al.~\cite{welch_frequency_2012}.

    We observe that the GM captures the frequency values of the translating coiling and alternating loops patterns remarkably well. However, the model is less accurate for the W pattern, and inaccurate for the meander pattern. For the W pattern, the GM predicts the frequency to decrease slightly with $V$, while the frequency remains unchanged in the experiments. For the meander pattern, which occurs at higher values of $V$, the GM captures the frequency variations poorly. We observed the frequency of the meander pattern to increase approximately linearly with $V$ in the experiments, but the GM predicts them to rise at an increasing rate. 
    
    \subsection{\label{Results: Amplitude}Amplitude}
\begin{figure*}[htbp]
	\centerline
        {\includegraphics[scale=0.65]{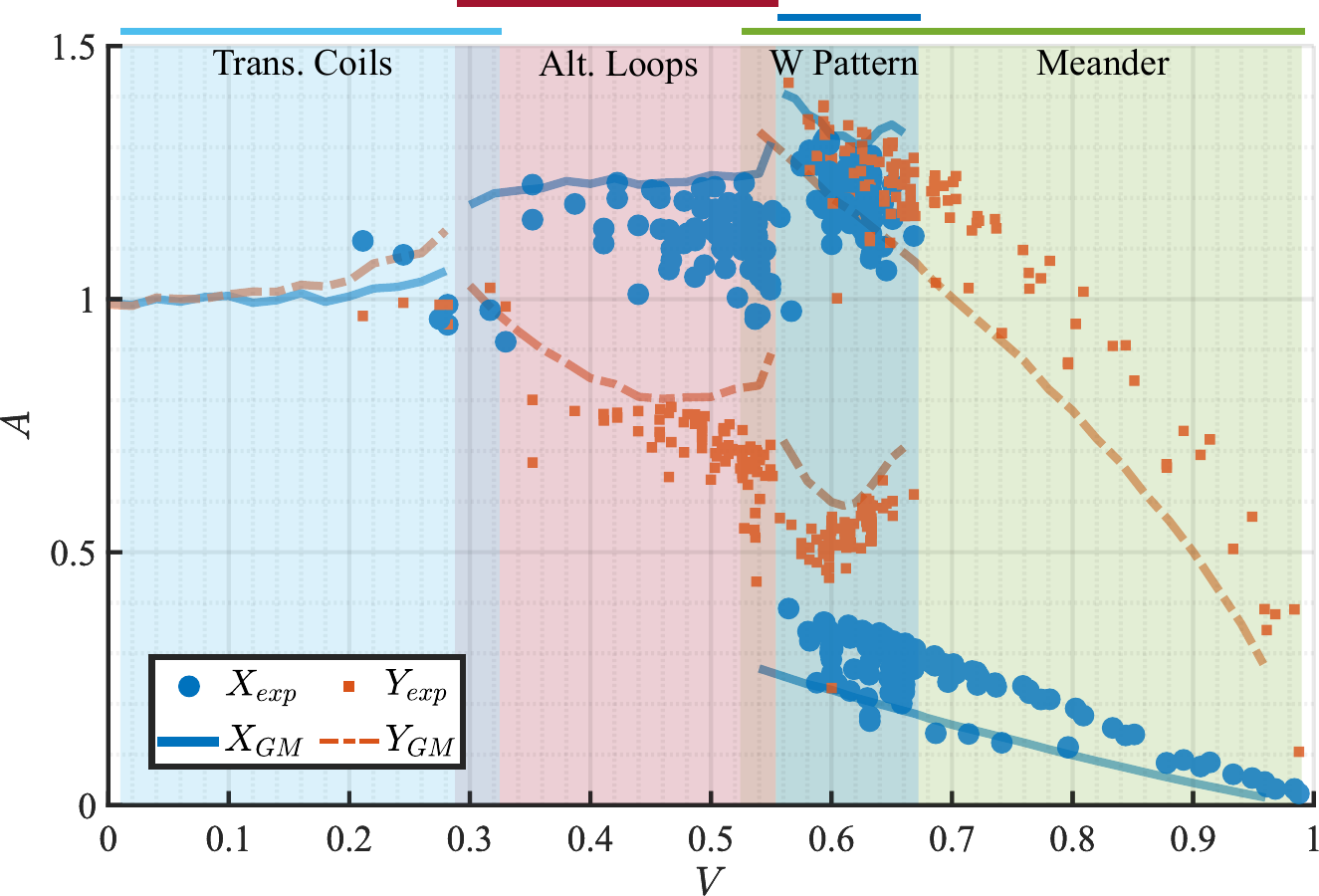}}
	\caption{(Color online) Dimensionless peak amplitudes from the Fourier spectrums of all collected patterns around the W region where $A = \hat{A}/\hat{R}_c$. The coloured horizontal lines above represent the extent of the predicted pattern regions over $V$.}
	\label{fig Amplitude}
    \end{figure*}

    The amplitudes $\hat{A}$ of the largest peaks in $x$ and $y$ from the frequency spectrum were extracted and compared to those obtained from the GM using the same methodology. Fig.~\ref{fig Amplitude} shows the non-dimensional amplitude, $A = \hat{A}/\hat{R}_c$, as a function of the non-dimensional belt speed, $V$. We observe that the scatter in the amplitude data is larger than the scatter in frequency; {\it cf.} Fig.~\ref{fig Freq}. This may be due to experimental errors. 
    
    In experimental data, the dominant amplitudes at low $V$ in both directions are around $A = 1$. As the pattern changes from translating coiling to alternating loops, the amplitudes separate with an upward jump in the $x$ direction and a downward jump in the $y$ direction. Both stay approximately constant for some $V$ until they decrease slightly before reaching the W pattern. When turning into a W pattern, the amplitude in $x$ further increases while the amplitude in $y$ decreases. As $V$ increases within the W region, $x$ peak decreases gradually while $y$ increases. Transitioning into meandering, the amplitude in the $x$ drops while the amplitude in $y$ becomes the highest. Both amplitudes decrease to zero where we find the straight pattern as $V$ approaches 1. 
    The trend of the $x$ and $y$ peaks as the pattern changes from translating coils to the W pattern can be interpreted as an elongation and a contraction of the orbit parallel and perpendicular, respectively, to the belt naturally due to the increase in belt speed. Although, it does not explain the trend within the patterns.

    The general trend from GM is somewhat similar to the experimental data throughout the entire $V$. At each transition, the amplitudes in the GM change similarly to experimental data. However, we note some clear differences. First, for translating coiling, alternating loops and W pattern, the amplitudes in GM are consistently higher than experimental data. For the meander, it is the opposite. 
    This amplitude shift may have resulted from tracking the thread position at some height above the belt in experiments. GM predicts the movement of the contact point on the belt. Since the thread curves to form the heel, the size of the orbit changes depending on the height at which the thread is tracked.
    In addition, within alternating loops, the $\hat{y}$ amplitude in GM is concave upwards while experimental data appears to be concave downwards. Finally, the amplitudes of the W pattern in the GM first decrease before increasing. The experimental data seems only to show a linear variation. The exact reasons for the differences remain unclear.

    \subsection{\label{Results: Period}Period}
    The period, $\hat{T}$, of each pattern was obtained by extracting the least common multiple of the fundamental period from the time series in the $x$ and $y$ directions. We non-dimensionalize the period with $\hat{T}_c$, the period of coiling on a stationary surface: $T = \hat{T}/\hat{T}_c$. Note that $\hat{T}_c$ is the inverse of $\hat{\Omega}_c$. 
    
\begin{figure*}[htbp]
	\centerline
        {\includegraphics[scale=0.65]{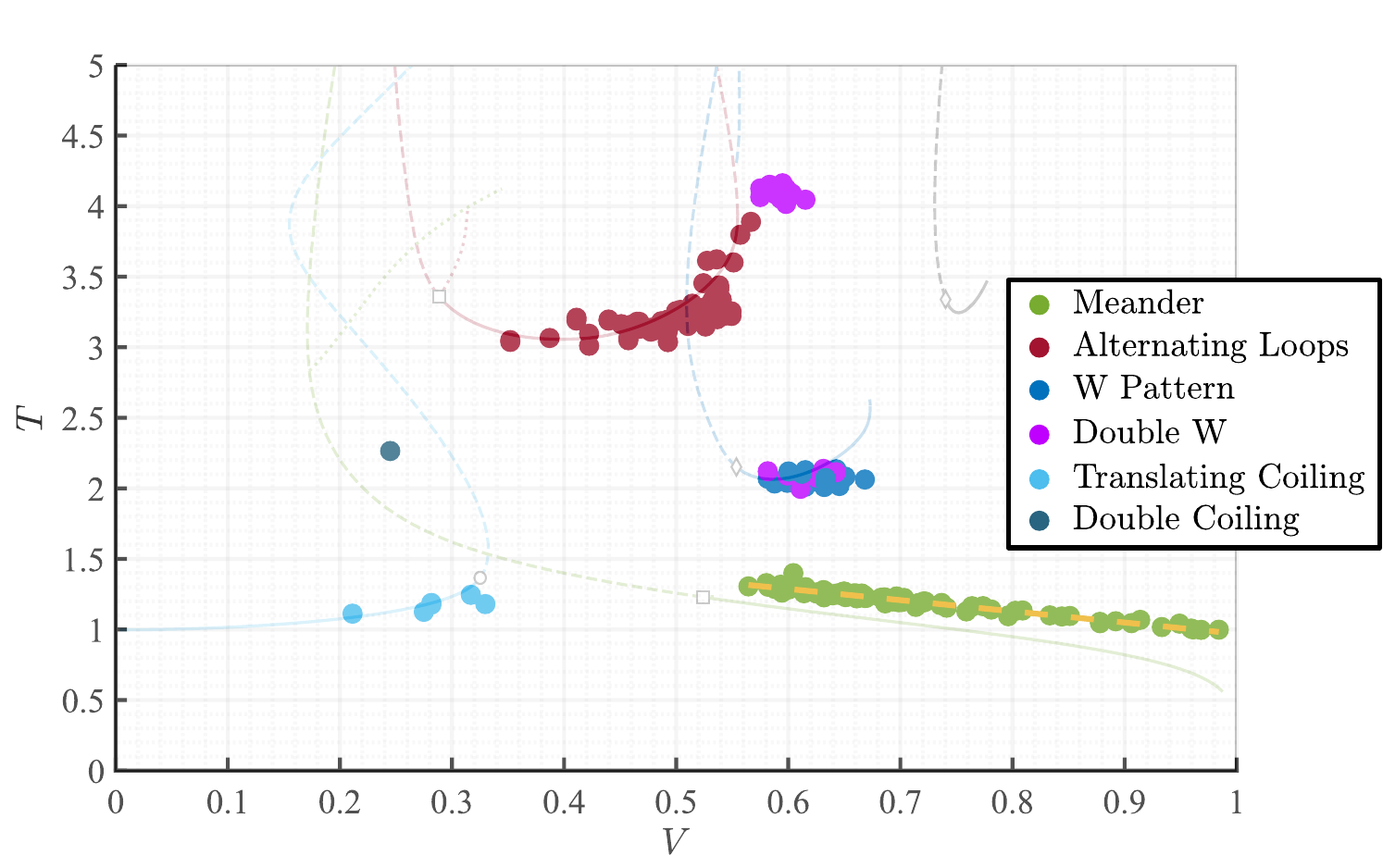}}
	\caption{(Color online) Dimensionless period of the patterns from experiments compared with the period variations of those from the GM. The dashed yellow line is a linear fit to the meander data. Diamond, square and circle markers indicate period-doubling, branch point and torus bifurcations, respectively. Unstable solutions are indicated using black dashed lines.}
	\label{fig Period}
    \end{figure*}

    Fig.~\ref{fig Period} shows $T$ as a function of non-dimensional belt speed, $V$. The period of the translating coiling pattern starts very close to $\hat{T}_c$, as expected from definition, and increases slowly with $V$. For the alternating loops pattern, it takes at least $3\hat{T}_c$ to complete a cycle. The period increases continuously with $V$ until a transition occurs to either a meander or a W pattern.
    The W pattern takes around $2\hat{T}_c$ to complete a cycle while the double W, as the name suggests, takes $4\hat{T}_c$. The period data clearly show that the double W pattern occurs at a lower value of $V$ than the W pattern. The periods of the two patterns remain roughly independent of $V$. The period of the meandering pattern decreases approximately linearly until it approaches $\hat{T}_c$ near the onset of the catenary (straight) pattern at $V=1$.

    Overall, the performance of the GM in predicting the period of the patterns is very similar to its performance in predicting the frequency; recall Fig.~\ref{fig Freq}. The predictions of the GM (curves in Fig.~\ref{fig Period}) follow the data closely for the translating coiling and alternating loops patterns. There is a slight discrepancy in the $V$-dependence of the W pattern: our data shows that the period of the W pattern does not depend on $V$. The period plot seems to suggest a connection between the alternating loops and the double W branch, though we were not able to capture the double W branch using the GM.  
    Throughout the entire range of $V$ occupied by the meander pattern, the GM predicts a lower period and decreases at a faster rate than the data. 

\section{\label{Disscussion Conclusion}Summary and conclusion}
We presented a detailed investigation of the patterns appearing at relatively low to moderate heights in a Fluid Mechanic Sewing Machine. In this height range, the patterns were expected to be well organized and governed solely by a simple theoretical model based on the geometry and local bending of the liquid rope at the contact point. This geometric model of coiling (GM) can predict the correct regions and sequence of the basic coiling patterns, but no detailed comparisons with the experimental results have been made until now. 

We built an experimental apparatus and collected new data to complement the existing results in the literature. We used the new data to review the basic patterns: translating coiling, alternating loops, W pattern and meander. We repeatably observed the W pattern in a specific region between the meandering and alternating loops pattern. This pattern was previously reported to occupy an insignificant space in the regime diagram. Our observation of the W pattern is consistent with the predictions of the GM. We observed the double W pattern, a period-doubled version of the W pattern. The double W pattern, coexisting in the same range of velocities as the W pattern, is reported for the first time, to the best of our knowledge. The double W pattern occurred near a period-doubling bifurcation point predicted by the GM. However, the observed pattern was different from the one predicted by the GM. 

By rescaling the data obtained into dimensionless parameters, we confirmed, in a limited region, that the sequence of the patterns depends on the ratio of belt speed to fluid speed and not the fall height. We found that the GM predictions follow the experimental observations well in many areas of the regime diagram. We also noted some differences, including some appearing in regions at very low heights. We suspect that the thread thickness during overlapping may hinder its trajectory making it more difficult to maintain the pattern. 
The most outstanding difference is in variations of the meandering frequency or period with the belt speed, which the GM replicates poorly. These deviations may be related to the limitation of the fitted curvature equation, $\kappa(r,\theta-\psi)$. 
The curvature equation was obtained based on data from the translating coiling pattern DVT~\cite{brun_GM_2015}, but captures the main characteristics of other patterns. This fit may deviate from the curvature of the meandering pattern, leading to the discrepancies observed in the frequency of the pattern. Unexpectedly, we also observed isolated instances of resonant patterns. These are patterns typically found at higher heights due to the stronger presence of inertia.

In conclusion, we have found that the dynamics governing the patterns at low to moderate heights may be more complex than previously thought. The GM, despite its apparent simplicity, captures the general traits of non-inertial patterns exceedingly well. However, the model can benefit from additional refinement to capture the more complex features observed in the experiments such as the period-doubled pattern. This could involve using the experimental data to update the curvature equation by traditional fitting or data-driven discovery techniques, or by incorporating missing physics in the computational techniques such as thread overlay. 
A global bifurcation analysis can further clarify the morphology of the patterns and the underlying transition dynamics between them. 
In summary, liquid coiling, even in the non-inertial regime, contains many unexpected and fascinating complexities that merit further experimental and theoretical investigation if we want to fully understand and predict the behaviour.

\begin{acknowledgments}
We thank S.W. Morris, University of Toronto, and P.T. Brun, Princeton University, for their insightful discussions of the results. We thank S.W Morris for sharing data from~\cite{welch_frequency_2012}. We thank B. Krauskopf, University of Auckland, for insightful comments about symmetries, underlying global bifurcations and suggestions for extending the bifurcation analysis. We thank H.K. Moffat, University of Cambridge, for a discussion of symmetries in the rotational FMSM. At Concordia, we extend our gratitude to Syn Furuli and Rafael Juarez Rodriguez, technicians, and Michael Rembacz, Engineer in Residence, for their invaluable input on the design and troubleshooting of our experimental apparatus. We acknowledge financial support from the Natural Sciences and Engineering Research Council of Canada through the Discovery Grant program.
\end{acknowledgments}

\bibliography{Ref}

\end{document}